\documentclass[11pt, conference, compsocconf]{IEEEtran}
\usepackage{times}
\usepackage{url}
\usepackage{latexsym}
\usepackage{vntex}
\usepackage[english]{babel}
\usepackage{graphicx}
\usepackage{amsmath}
\usepackage{amsfonts}
\usepackage{framed}
\usepackage{multirow}

\begin{document}

\title{A Vietnamese Information Retrieval System for Product-Price}

\author{\textbf{Tien-Thanh Vu} and \textbf{Dat Quoc Nguyen}\\
Faculty of Information Technology\\
University of Engineering and Technology \\
Vietnam National University, Hanoi\\ 
\{tienthanh\_dhcn, datnq\}@vnu.edu.vn
}

\maketitle
\thispagestyle{empty}

\begin{abstract}
  A price information retrieval (IR) system allows users to search and view differences among prices of specific products. Building product-price driven IR system is a challenging and active research area. Approaches entirely depending products information provided by shops via interface environment encounter limitations of database. While automatic systems specifically require product names and commercial websites for their input. For both paradigms, approaches of building product-price IR system for Vietnamese are still very limited. In this paper, we introduce an automatic Vietnamese IR system for product-price by identifying and storing  Xpath patterns to extract prices of products from commercial websites. Experiments of our system show promising results.
  
\end{abstract}


%
\IEEEpeerreviewmaketitle

\section{Introduction}

A price information retrieval (IR) system allows users to search and view differences between prices of specific products. The system mainly focuses on collecting and updating price information of products crawled from commercial websites. There are generally two main approaches to build a product-price IR system: \medskip

\begin{itemize}
\item The first bases on interaction between shops and the product-price IR system, in which the system  creates an interface environment allowing shops to directly provide product-price information to system. This system type encounters limitations of database in entire dependence on the shops. Because the price always changes over time, it requires price information to be constantly updated to the database.\medskip
 
\item The other automatically updates the IR system's database by crawling on commercial websites of shops to extract product-price information. However, this system type requires that product names must be firstly provided and commercial websites must be specified.\medskip
\end{itemize}

In this paper, we introduce a price-driven Vietnamese IR system for products in handling above mentioned drawbacks. With a small number of initial seed product names, our system's front-end component automatically identifies related commercial websites and corresponding Xpath patterns. Then the back-end component uses the related websites and Xpath patterns to collect and update the database of names and prices from crawled products. \medskip

The rest of paper is organized as follows: in section \ref{sec:relatedworks}, we
provide some related works. We describe our system and our experiments in section \ref{sec:oursystem} and section \ref{sec:experiment} respectively. The conclusion and future works will be presented in section \ref{sec:conclusion}.

\section{Related works}
\label{sec:relatedworks}

\begin{figure*}[ht]
\centering
\includegraphics[width=14cm]{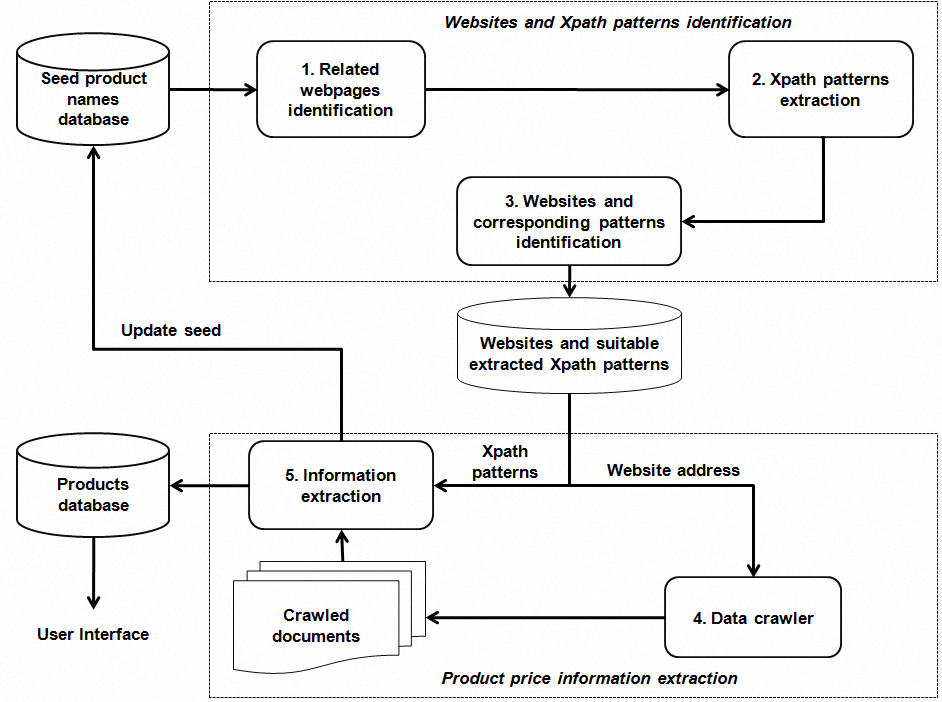}
\caption{Architecture of our price IR system.}
\label{Figure1}
\end{figure*}

There have already existed numerous shopping search engines, but they mostly require product-information to be collected and updated manually. PriceScan%
\footnote{%
	www.http://www.pricescan.com
} and GoogleProduct%
\footnote{%
	http://www.google.com/prdhp
}  show products from a manually updated database. Kelkoo%
\footnote{%
	http://www.kelkoo.co.uk
}  and Yahoo! Shopping%
\footnote{%
	http://shopping.yahoo.com
}  utilize database frameworks where merchants submit their products to be manually classified according to a defined structure. Recently, some Vietnamese shopping search engines have been presented such as: www.vatgia.com, www.aha.vn. But all of them is built according to the first main approach shown in the introduction.

The related works to our approach come from primary field of information extraction from semi-structured webpages. Kushmerick et al \cite{kushmerick97wrapper}, Muslea et al.\cite{Muslea}, Freitag and Kushmerick \cite{freitag00boosted}, Cohen et al.\cite{Cohen02} introduced and improved wrapper induction method which generates extraction rules in using machine learning approach. From a few training webpages which manually predetermine the target-items, the method learns to extract rules. The rules then are applied to detect target-items from other pages. 

Nguyen et al. \cite{NguyenNPB09}  proposed an approach to automatically extract primary text content of webpages by identifying and storing templates representing the Xpath structure of text content blocks in websites. Carlson and Schafer \cite{CarlsonS08} described bootstrapping information extraction method which only annotates 2–5 webpages over 4–6 websites. The obtained results significantly outperform the baseline approach with the extraction accuracy of 83.8\% on job offer websites and 91.1\% on vacation rental websites. Crescenzi et al. \cite{Crescenzi01} presented Roadrunner system which automatically extract information by comparing structure of web pages in requirement of extracted data to be labelled by user.

Zhang et al.\cite{Zang09} described an ontology-based e-commerce product information retrieval framework and presented an ontology-based adaptation of the classical Vector Space Model in considering the weight of product's attributes.

\section{Our Vietnamese IR system for product-price}
\label{sec:oursystem}
In this section, we describe our product-price information retrieval system . Figure  \ref{Figure1} shows our price IR system's architecture. Our system contains two components front-end and back-end. The front-end takes input of seed product names to automatically identify suitable websites and  Xpath patterns. The back-end component of product-price information extraction crawls data from URLs in the suitable websites and uses Xpath patterns to extract names and prices information of products in crawled data. The extracted information will be updated into databases of products and seed product names.

\subsection{The front-end component of websites and Xpath patterns identification}
The font-end component consists  of three modules of ``related webpages identification'', ``Xpath patterns extraction'', and ``websites and corresponding patterns identification''.\medskip 

\subsubsection{Related webpages identification module}
This module takes a set of seed product names as the input and returns webpages relating to the product names.\medskip

Based on specific characteristics of commercial websites, we create particular queries matching product names to Google search engine by utilizing some defined templates.
For example: instead of using query \textit{``ipad 2''}, the query \textit{``ipad 2'' + ``vnđ or usd''} is automatically generated in the use of template \textit{\textbf{``product\_name''} + ``vnđ or usd''}, and it is sent to Google search engine. All top five webpages of returned results by the Google are from commercial domains.\medskip

\subsubsection{Xpath patterns extraction module}
The input of this module is a product name and a related webpage returned by Google search engine. The output is actual price and Xpath patterns to be used to detect product names and the actual prices. \medskip

\begin{figure}[ht]
\centering
\includegraphics[width=7cm]{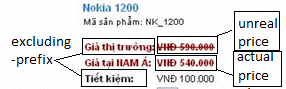}
\caption{An example of actual price extraction.}
\label{fig:nokia1200}
\end{figure}

For example, with given product name of \textit{``Nokia 1200''} and one of related webpages identified from the previous module, the patterns extraction module returns results of  \textit{``VNĐ 540.000''} (figure \ref{fig:nokia1200}) and Xpath patterns of \textit{``HTML $\rightarrow$ BODY $\rightarrow$ TABLE[1] $\rightarrow$ TR[1] $\rightarrow$ TD[1] $\rightarrow$ \textbf{product\_name}''} and \textit{``HTML $\rightarrow$ BODY $\rightarrow$ TABLE[1] $\rightarrow$ TR[2] $\rightarrow$ TD[2] $\rightarrow$ \textbf{actual\_product\_price}''}.
\medskip

 Because webpages on the same website usually have similar structures, we can use these Xpath patterns to extract product names and  corresponding actual prices from other webpages.

The Xpath patterns extraction module has 2 sub-modules:
\begin{itemize}
\item \textbf{The first sub-module} identifies leaf node in Document Object Model (DOM) tree corresponding with HTML source code of the input related webpage, in which the node contains the text string matching the input product name. The first sub-module generates Xpath pattern by using traversal path from root node of DOM tree to the detected leaf node. 

\item \medskip\textbf{The second sub-module} firstly find the leaf node in the DOM tree in which the node contains text string of actual price, and then the second generates corresponding Xpath pattern. The module detects the node containing text string catching ``actual price'' through following steps:
\end{itemize}

\begin{itemize}
\item[-] \textit{Step 1}: Detect all text strings representing numbers  in the input webpage by employing  basic regular expressions. For example, in figure \ref{fig:nokia1200}, extracted text strings are \textit{``1200''}, \textit{``590.000''}, \textit{``540.000''} and \textit{``100.000''}.

\item[-] \medskip \textit{Step 2}: From extracted text strings via step 1, the module identifies all text strings describing maybe-actual prices through prefix, suffix,  and excluding rules:

\medskip \textit{Prefix rule}: A number represents a product-price if it is preceded by ``Giá$_{price}$'' or  ``VNĐ$_{Vietnam\ dong}$'',...

\medskip \textit{Suffix rule}: A number represents a product-price if is followed by ``VNĐ$_{Vietnam\ dong}$'', ``USD'', ``Đ$_{dong}$'', ``\$'',...

\medskip \textit{Excluding rules}: A text string does not represent an actual price if it is preceded by ``Giá cũ$_{Old\ price}$'', ``Giá thị trường$_{Market\ price}$'', ``Tiết kiệm$_{Save}$'',... A text string does not represent an actual price if it is stored by DOM tree nodes of tags <strike> or <s>. For example, in figure \ref{fig:nokia1200}, text string \textit{``VNĐ 590.000''} is  not actual price because the text string belongs to tree node of tag <strike>. Text string \textit{``VNĐ 100.000''} followed by ``Tiết kiệm$_{Save}$'' is not actual price.

\item[-] \medskip\textit{Step 3}: Determine the actual price if there are some maybe-actual prices. It needs to  examine relationship between name and actual price of product. The relationship means that product's name and product's actual price are held by two closet nodes of DOM tree. It is a specific characteristic of commercial webpages.

\medskip For example: with the Xpath pattern \textit{HTML $\rightarrow$ BODY $\rightarrow$ TABLE $\rightarrow$ TR $\rightarrow$ TD $\rightarrow$ DIV[1]   $\rightarrow$  \textbf{product\_name}} generated from the first sub-module to extract the input product name, and a Xpath pattern corresponding with a maybe-actual price \textit{HTML $\rightarrow$ BODY $\rightarrow$ TABLE $\rightarrow$ TR $\rightarrow$ TD $\rightarrow$ DIV[2] $\rightarrow$ FONT $\rightarrow$ \textbf{product\_price}}.
The similar-measure is 5 overlap steps \textit{HTML[1] $\rightarrow$ BODY[2] $\rightarrow$ TABLE [3]$\rightarrow$ TR[4] $\rightarrow$ TD[5]}.
The Xpath pattern to extract price, that has highest similar-measure in comparison with the Xpath pattern used to extract input product name, is selected as output pattern to extract actual price.

\end{itemize}
\medskip

\subsubsection{Websites and corresponding patterns identification module}
This module returns commercial websites and suitable Xpath patterns to be used to generate names and actual prices of products from the themselves. The module counts number webpages from each website in which the webpages have same identified Xpath patterns determined the previous module. If the number is greater than a given threshold, the website is considered as a commercial website and the corresponding Xpath patterns are suitable patterns.  

\subsection{The back-end component of product-price information extraction}
In this component, we focus on two modules Data crawler and Information extraction. The component takes front-end's output as input of identified commercial websites and suitable Xpath patterns matching with each website. HTML documents from  the websites will be collected in the use of Data crawler module via browsing hyper-links in each crawled document. \medskip

The information extraction module uses the input of collected HTML documents and suitable Xpath patterns to extract information of product names and actual product-prices. Extracted information then will be updated into Products database and Seed product names database (figure \ref{Figure1}).

\section{Experiments}
\label{sec:experiment}

We built our system on computer of Intel Celeron@CPU 2.66GHz and RAM 768MB. With initial set of 334 seed product names from many product types such as mobile phone, computer, camera, jewellery, household items,... in 30 hours, our system collected 47856 products from 125 determined commercial websites in which 34012 products are unique. For example, ``Lenovo ThinkPad T61'' and ``IBM T61'' are considered as the same one while ``Nokia 1200 black'' and ``Nokia 1200 white'' are different.
In order to clearly evaluate our system's modules, we present some experiments as follows. 

\subsection{Experiment of ``Related webpages identification''}

To evaluate the template \textit{``\textbf{product\_name}'' + ``VNĐ or USD''} that we employed to create queries, we randomly selected products of ``Nokia 1200'', ``Lenovo Thinkpad t61'' and ``Canon PowerShot G10''. Table \ref{tab:webIdenMod} shows the number of commercial webpages containing product name and its actual price, in top 10, 30 and 100 returned webpages by using Google Search Engine. Other returned results by Google belong to webpages of news, forums,...

\begin{table}[ht]
\caption{Number of commercial webpages returned by Google Search Engine}
\label{tab:webIdenMod}
\centering
\begin{tabular}{ p{2cm} | p{3cm} | p{2.5cm} }
\hline
Product name & Number of related webpages by Google & Number of commercial webpages \\
\hline
\multirow{3}{*}{Nokia 1200} & 10 & 8 \\ \cline{2-3}
 & 30 & 23 \\\cline{2-3}
 & 100 & 68 \\ \hline
\multirow{3}{2cm}{Lenovo Thinkpad t61} & 10 & 10 \\ \cline{2-3}
 & 30 & 23 \\\cline{2-3}
 & 100 & 43 \\ \hline
 \multirow{3}{2cm}{Canon PowerShot G10} & 10 & 9 \\ \cline{2-3}
 & 30 & 19 \\\cline{2-3}
 & 100 & 45 \\
\hline
\end{tabular}
\end{table}

\subsection{Experiment of actual price extraction in ``Xpath patterns extraction'' module}

To right examine extraction-ability of this module, we used the commercial webpages determined in the previous experiment (table \ref{tab:webIdenMod}). In this experiment, we consider F$_{measure}$ as a metric to evaluate the accuracy of price extraction as presented in table \ref{tab:accuracy}.

\medskip F$_{measure}$ = $\dfrac{2 * Recall * Precision}{Recall + Precision}$

\medskip \textit{Precision} is defined as the ratio between the number of extracted actual-prices and the total number of detected prices, while  \textit{Recall} is defined as the ratio between the number of extracted actual-prices and the actual number of actual-prices. 

\setcounter{table}{3}
\begin{table*}[ht]
\caption{Accuracy of product's name and price extraction}
\label{tab:nameprice}
\centering
\begin{tabular}{p{4.5cm} p{2.75cm}   p{2.5cm} p{5.5cm}}
\hline Website & Number of crawled webpages & Number of commercial webpages & Number of pairs of extracted product name and corresponding actual price\\ 
\hline
www.dienthoaididong.com.vn & 850 & 792 & 743 (93.81 \%)\\
\hline
www.trananh.vn & 800 & 711 & 416 (58.5 \%)\\
\hline
\end{tabular}
\end{table*}

\setcounter{table}{2}
\begin{table}[ht]
\caption{The accuracy of price extraction}
\label{tab:accuracy}
\centering
\begin{tabular}{ p{2cm} | l l l}
\hline
Product name & Recall & Precision & F-measure \\
\hline
\multirow{3}{*}{Nokia 1200} & 8/8 (1.0) & 8/8 (1.0) & 100 \% \\ \cline{2-4}
 & 23/23 (1.0) & 23/26 (0.88) & 93.88 \% \\\cline{2-4}
 & 67/68 (0.99) & 67/70 (0.96) & 97.10 \% \\ \hline
\multirow{3}{2cm}{Lenovo Thinkpad t61} & 9/10 (0.9) & 9/10 (0.9) & 90 \% \\ \cline{2-4}
 & 22/23 (0.96) & 22/25 (0.88) & 91.67 \% \\\cline{2-4}
 & 40/43 (0.93) & 40/46 (0.87) & 89.89 \% \\ \hline
 \multirow{3}{2cm}{Canon PowerShot G10} & 9/9 (1.0) & 9/9 (1.0) & 100 \% \\ \cline{2-4}
 & 18/19 (0.95) & 18/21 (0.86) & 90 \%\\\cline{2-4}
 & 44/45 (0.98) & 44/50 (0.88) & 92.63 \%\\
\hline
\end{tabular}
\end{table}

\begin{table}[ht]
\caption{Accuracy of commercial websites identification}
\label{tab:websiteInden}
\centering
\begin{tabular}{p{2cm}|  l |l}
\hline Top results of Google & Identified websites & Accuracy\\
\hline 
\multirow{4}{*}{10} & www.123mua.com.vn & \multirow{4}{*}{100 \%} \\ 
 & www.vatgia.com & \\
  & www.vinacms.vn &  \\
 & www.chodientu.vn &  \\ \hline
\multirow{9}{*}{100} & www.123mua.com.vn & \multirow{9}{*}{100 \%} \\ 
 & www.vatgia.com & \\
  & www.vinacms.vn &  \\
 & www.chodientu.vn &  \\
  &www.enbac.com & \\
  & www.quangcaosanpham.com &  \\
 & www.aha.vn &  \\
  & www.dienthoaididong.com.vn & \\
  & www.trananh.vn &  \\
\hline
 
\end{tabular}
\end{table}

\subsection{Experiment of ``commercial websites identification''}
For initial set of 4 products of ``Nokia 1200'', ``Nokia e71 white steel'', ``Nokia 1202'' and ``Nokia 6300 silver'' and a defined threshold of 3 to determine commercial websites, table \ref{tab:websiteInden} gives accuracy of 100\% for the first component on both cases of taking top 10 and 100 related webpages returned by Google in the first module of our system.

\subsection{Experiment of ``information extraction'' module}

This experiment shows our evaluation in the use of identified Xpath patterns to extract names and prices of products. 
From the output of the front-end component in taking the set of 4 products as input that is described in the ``commercial websites identification'' experiment, we selected two websites \textit{www.dienthoaididong.com.vn} and \textit{www.trananh.vn} and their corresponding suitable Xpath patterns to perform the evaluation. \medskip

We randomly crawled a number of webpages per each selected website by ``Data crawler'' module, in which there are many webpages coming from website's news and forum. We only calculated the accuracy based on number of commercial webpages. Table \ref{tab:nameprice} presents promising results that the information extraction module well performed on the website www.dienthoaididong.com.vn. 
The website www.trananh.vn has different Xpath structures for representing different product categories such as computer, camera, household items,... in HTML documents, therefore, with 4 given seed product names only belonging to the category of mobile phones, 416 extracted products from  www.trananh.vn only belong to the mobile phone category. Consequently, the returned result is not high.  It is easy to improve the result by taking seed products from all kinds of categories.

\section{Conclusion}
\label{sec:conclusion}
We believe on fast scalability of our system. Our system can identify more sites and Xpath patterns depending on the number of initial seed product names. Because extracted product names returned by information extraction module always are updated into the seed products database, the database always is expanded. In addition, it is possible for our proposed system's architecture to adapt to a new language by changing the rules according to the new one.  \medskip

In this paper, we introduce an automatic product-price information retrieval system for Vietnamese commercial sites. With a small number of seed product names, our system automatically detects commercial sites, generates corresponding Xpath patterns. Our  system then uses identified information to extract name and actual price of crawled products. \medskip

The experiment results are promising; with 334 initial product names, our system determined 125 commercial sites and collected 47.856 products in 30 hours. In the future, we will extend our system's rules driving to collect information of size, weight, guarantee period, and other features of products. 

\section*{Acknowledgement}

The authors would like to acknowledge Vietnam National Foundation for Science and Technology Development (NAFOSTED) for their financial support to present the work at the conference.

\bibliographystyle{IEEEtran}
\bibliography{refers}

\end{document}